# Air-stable lithiation engineering of MoS$_2$ for direct-bandgap multilayers

Qi Fu[1,2,3,9], Yichi Zhang[1,2,3,9], Jichuang Shen[4,5], Siyuan Hong[6], Jie Wang[7], Chen Wang[1,2,3], Jingyi Shen[1,2,3], Wei Kong[4], Guolin Zheng[7], Jun Yan[8], Jie Wu[2,3*], Changxi Zheng[2,3*]

[1]School of Physics, Zhejiang University, Hangzhou, 310027, China.
[2]Research Center for Industries of the Future, Westlake University, Hangzhou 310024, China.
[3]Department of Physics, School of Science, Westlake University, Hangzhou 310024, China.
[4]School of Engineering, Westlake University, Hangzhou 310030, China.
[5]School of Materials Science and Engineering, Zhejiang University, Hangzhou, China.
[6]Instrumentation and Service Center for Physical Sciences, Westlake University, Hangzhou 310024, Zhejiang Province, China.
[7]Anhui Provincial Key Laboratory of Low-Energy Quantum Materials and Devices, High Magnetic Field Laboratory, HFIPS, Chinese Academy of Sciences, Hefei, Anhui 230031, China.
[8]Department of Physics, University of Massachusetts Amherst, Amherst, Massachusetts 01003, USA.
[9]These authors contributed equally: Qi Fu, Yichi Zhang.
Email: zhengchangxi@westlake.edu.cn; wujie@westlake.edu.cn

**Abstract**

Due to its sizable direct bandgap and strong light-matter interactions, the preparation of monolayer $MoS_2$ has attracted significant attention and intensive research efforts. However, multilayer $MoS_2$ is largely overlooked because of its optically inactive indirect bandgap caused by interlayer coupling. It is highly desirable to modulate and decrease the interlayer coupling so that each layer in multilayer $MoS_2$ can exhibit a monolayer-like direct-gap behavior. Here, we demonstrate the nanoprobe fabrication of $Li_xMoS_2$-based multilayers exhibiting a direct bandgap and strong photoluminescence emission from tightly bound excitons and trions. The fabrication is facilitated by our newly developed Li-ion platform, featuring tip-induced Li intercalation, air stability and rewritability. Raman characterizations reveal that controlled Li intercalation effectively transforms multilayer $MoS_2$ into the stack of multiple monolayers, leading to a 26-fold enhancement of photoluminescence, compared to a monolayer. This intercalation result is different from existing observations of transforming $MoS_2$ multilayers into metallic phases.

**Main**

van der Waals (vdW) transition metal dichalcogenides (TMDCs), such as $MoS_2$, $WS_2$, and $WSe_2$, undergo a remarkable transition from an indirect to a direct bandgap when thinned down to monolayer thickness[1, 2, 3]. This direct bandgap is a pivotal property for the intensive interest in the field of two dimensional (2D) materials, as it enables strong light-matter interactions and intense photoluminescence (PL)[4]. Owing to the enhanced Coulomb interactions and spin-orbital coupling, the PL of monolayer TMDCs is characterized by tightly bound excitons and trions, as well as a valley degree of freedom, which are essential for exploring quantum phenomena and the development of next-generation optoelectronics[5, 6, 7, 8, 9].

To this end, the community of 2D materials has primarily focused on developing technologies for preparing monolayer TMDCs over the years. These technologies include mechanical exfoliation, chemical vapor deposition (CVD), chemical exfoliation, and laser ablation[10, 11, 12, 13, 14, 15, 16]. Meanwhile, recent studies have demonstrated intense PL in bulk TMDCs after the intercalation of large organic molecules into their vdW gaps[17, 18]. Given these findings, it is of great importance to explore whether intense PL can be achieved in Li-intercalated $MoS_2$ multilayers, considering the small size of Li ions for fast interlayer motion and the broad applications of Li ion-based technology[19, 20, 21]. Unfortunately, based on extensive experimental characterizations from Li-assisted chemical exfoliation and Li ionic gating, it appears that intense PL and Li intercalation are incompatible[22, 23]. This incompatibility arises because Li intercalation introduces a significant amount of n-type doping, which quenches PL and can even transform semiconducting TMDCs into the metallic 1T phase at substantial concentration[24, 25, 26].

In this article, we report the achievement of intense PL composed of excitons and trions in multilayer $MoS_2$ after Li intercalation. For clarity, we denote Li intercalated $MoS_2$ multilayer as $Li_xMoS_2$ from here on. The Li intercalation is performed on our newly developed Li ionic platform, which consists of exfoliated $MoS_2$ thin flakes deposited on a solid ionic substrate and covered by a 27-nm $Al_2O_3$ thin film. Unlike prevalent solution-based doping technologies[10, 14, 15, 22, 26], here the Li intercalation is achieved in a dry environment. Particularly, the Li intercalation can be locally controlled for doping patterning by an atomic force microscopy (AFM) cantilever, using a bias voltage applied either to the backside of the substrate or to the cantilever itself, without involving any electrolytes or chemicals. Based on combined characterizations using AFM, PL, and Raman spectroscopy, the strong PL in multilayer $Li_xMoS_2$ is attributed to the isolation of interlayer coupling by Li intercalation and the reduction of Li-induced n-type doping by the ionic substrate beneath after its Li depletion. In addition to strong PL, our unique Li ionic platform features on-demand doping patterns with submicrometer resolution, tunable carrier density, air stability and rewritability. Our results not only provide a unique platform for material fabrication and the development of doping patterned electronics, but also further the understanding of Li interactions with van der Waals (vdW) materials.

**Design concept of a Li ionic platform for doping patterning using nanoprobe**

Figure 1a presents a schematic illustration of the structure and setup for Li dopant patterning and visualization by using AFM and its accompanied techniques including piezoresponse force microscopy (PFM) and Kelvin probe force microscopy (KPFM)[27, 28, 29]. As shown, the technique has a simple structure where a 2D material (MoS$_2$) is deposited on a Lithium-Ion Conducting Glass-Ceramics (LICGC) substrate and covered by a layer of oxide dielectrics[30]. Such kind of structure is denoted as Al$_2$O$_3$/MoS$_2$/LICGC from here on. Throughout the scanning procedure, either the reverse side of LICGC or the AFM cantilever is electrically grounded. Both work equally well.

The detailed process for doping patterning in arbitrary region is given in Fig. 1b. Firstly, the channel region of interest on the prefabricated 2D device is determined by AFM scanning, and the original electronic property of this region is observed using KPFM. Secondly, doping patterns are rendered into the selected region by translating the designed voltage pattern of AFM tip to the sample surface using PFM lithography technique that has been widely used in ferroelectric materials[31]. Afterwards, the doping patterns can be confirmed by KPFM visualization.

Figure 1c presents a KPFM image of a MoS$_2$ device with Ti/Au contacts covered by a 27 nm thick alumina (Al$_2$O$_3$) film. The optical microscopy image of the device is displayed in the Inset. Raman and photoluminescence (PL) spectra confirm the monolayer thickness of the MoS$_2$ film[1, 2, 32], which was grown by CVD, see Methods and Supplementary Fig. 1. Following the growth, the MoS$_2$ film is carefully transferred onto the LICGC substrate using a wet chemical process. Afterward, the device is fabricated and encapsulated with an Al$_2$O$_3$ film via atomic layer deposition (ALD), as described in the Methods section. Given the exceptional dielectric properties of Al$_2$O$_3$, the overlaying thin film does not hinder the long-range Coulombic interactions between the AFM tip and the device, thus enabling the capture of a high-resolution KPFM image of the device[33]. The surface potential of the Ti/Au contact, as shown in Fig. 1c, is approximately 200 mV lower than that of monolayer MoS$_2$. This implies that the Fermi level of the Ti/Au contact is positioned below that of MoS$_2$. Consequently, the work function of MoS$_2$ is smaller than that of Ti/Au, aligning with the findings reported in the literature[34]. It should be noted that the surface of LICGC is not well grounded during KPFM measurements due to its electrical insulation, resulting in a floating potential that hinders the precise measurement of the work function of the device on top. However, the relative surface potential differences between different components on the surface, such as the Ti/Au contacts and the MoS$_2$ channel here, can be measured accurately. Prior to measuring the samples, the work function of each AFM cantilever is calibrated using an Au-Si-Al sample, as shown in Supplementary Fig. 2.

Since the work function difference at the interface buried by the Al$_2$O$_3$ film can be clearly imaged using KPFM, we further explore the capability of moving Li ions in LICGC substrate by applying voltage to the conductive AFM tip. For this purpose, PFM lithography is employed, as it has been widely used to write arbitrarily designed voltage patterns onto ferroelectric surfaces[31]. For comparative analysis, PFM lithography is performed on both the bare LICGC substrate and the one covered by a 27 nm Al$_2$O$_3$ film. Note that LICGC covered by an $Al_2O_3$ film without $MoS_2$ is denoted as $Al_2O_3$/LICGC from here on. A voltage pattern of a Quick Response (QR) code, as shown in Supplementary Fig. 3a, is written onto both types of samples. Fig. 1d displays the KPFM image of the surface of LICGC covered by the $Al_2O_3$ film after writing the QR code, and the corresponding AFM topography is shown in Supplementary Fig. 3b. The contrast observed within the KPFM image corresponds to the distinct doping regions which are precisely defined by the PFM tip voltage. In contrast, no distinct KPFM contrast is observed on the surface of the bare LICGC, as seen in Supplementary Fig. 4. The absence of contrast is likely due to the reactivity of Li ions with ambient air components, whereas the $Al_2O_3$ film can effectively protect them from such interactions.

**Parameters for doping patterning and electronic behaviors**

To further explore the PFM doping mechanism in $Al_2O_3$/$MoS_2$/LICGC, doping arrays were generated by varying the tip voltage in a series from 1 V to 10 V, with steps of 1 V, see Fig. 2a and 2b. The corresponding line profiles of the dopant rows defined by the tip voltage series are presented in Fig. 2c. As shown, the electron doping level increases as the tip voltage rises. By measuring the peak values of potential differences as a function of tip voltage, a linear relationship between the surface potential shift and the tip voltage is obtained (see Fig. 2d). It should be noted that as the tip voltage increases, the peak value of surface potential rises, while the valley value decreases, as observed in Supplementary Fig. 5. These results suggest that the motion of Li ions, induced by the tip voltage, is not only in the direction of substrate thickness but also along the surface, due to the sharpness of the AFM tip. Thus, the accumulation of Li ions is usually accompanied by Li depletion in the vicinity, generating reverse doping. Fig. 2e presents the doping patterns, including a chessboard and fine bars, created by alternating the tip voltage from positive to negative, see Supplementary Fig. 6 for a PFM voltage pattern. By taking the line profile of the fine bars (Fig. 2f), the spatial resolution of the doping pattern, which is around 517 nm, can be determined. All results indicate that the doping level, as well as the features and size of the doping pattern, are highly controllable in our solid-state ionic chip.

Further confirmation of tip-writing controlled doping can be obtained through current-voltage (*I-V*) curve measurements of $Al_2O_3$/$MoS_2$/LICGC devices. As shown in Fig. 2g, the $MoS_2$ is electron-doped after being deposited on the LICGC substrate due to the presence of Li ions on the substrate surface and the native doping induced by CVD growth process. However, the *I-V* curve exhibits Schottky diode-like characteristics after PFM writing with a negative tip voltage (Fig. 2h). The Schottky diode-like *I-V* curve is caused by the work function difference between the lightly doped $MoS_2$ monolayer and the Ti/Au electrodes, as well as the asymmetric contact areas of the source and drain electrodes, which is due to the precision limits of general microfabrication. After further writing using higher negative voltage, the $Al_2O_3$/$MoS_2$/LICGC device eventually exhibits insulator-like properties (Fig. 2i). The results of electronic devices suggest that negative tip voltage would expel Li ions away from $MoS_2$ film and remove electron dopant. Moreover, as shown in Supplementary Fig. 7, the tip induced doping is air stable due to the protection of $Al_2O_3$ thin film[35].

**Rewritable doping patterns on demand**

Thus far, our results demonstrate that tip voltage patterning is an effective method for fabricating electronic devices with precise control over doping levels and regions. Fig. 3a, g presents KPFM images depicting a logo created by doping patterning, utilizing 8 V and -8 V tip voltages, respectively. The corresponding voltage maps are shown in Supplementary Fig. 8. As observed, the uniform contrast in the images indicates that a consistent level of doping has been achieved. After applying -3.4 V to the top row and 2.5 V to the bottom row regions, the doping patterns are nearly erased, leaving only a faint signal, as shown in Fig. 3b, h. This process is repeatable, and here we illustrate its repeatability over three cycles, as visualized by KPFM in Fig. 3c-f and i-l. It should be noted though, that there are residual signals which intensify with each successive cycle.

The exceptional repeatability of our solid-state ionic chip paves the way for the fabrication of memory electronics. As depicted in Fig. 3m, both high and low resistance states can be induced through tip voltage writing, resulting in an order of magnitude change in channel current. The device's performance is reproducible, and Fig. 3n demonstrates multiple consecutive repetitions here.

**Nanoprobe fabrication of multilayer Li$_x$MoS$_2$ with a direct bandgap**

Li-ion doping, in contrast to traditional ion implantation for dopant patterning, modifies not only the Fermi level of the material but also has the potential to decouple the interlayer interactions of vdW materials through Li intercalation[24]. A prime example is the chemical exfoliation of vdW materials facilitated by Li intercalation[14, 15]. This ionic doping characteristic differs from the electronic modulation effects observed in 2D electronics due to electrostatic gating[36].

Fig. 4a displays the optical microscope image of two mechanically-exfoliated MoS$_2$ thin flakes, which have been deposited on a LICGC substrate and coated with a 27 nm Al$_2$O$_3$ film via ALD (refer to the Methods section). Following PFM writing at 10 V to the two flakes, their optical contrast becomes weaker and even vanishes at edge regions, as shown in Fig. 4b. The change in optical contrast is a strong indicator of Li intercalation, which normally starts from edges and progresses towards the center, as observed by other groups[10, 19, 26]. The AFM topography of the flakes after Li intercalation is presented in Fig. 4c. The surface height variations between the edges (transparent) and the central regions (non-transparent) are attributed to Li intercalation, as shown in Supplementary Fig. 9. However, precise measurement of the height difference is challenging due to the roughness of the LICGC surface.

Raman and PL mappings were carried out to investigate the two flakes (see Fig. 4 and Methods for more details). Fig. 4c, d show the intensity mappings of the intra- and inter-layer Raman modes, located at 408 cm$^{-1}$ and 31 cm$^{-1}$, respectively[37]. It should be noted that these optical characterizations were performed in air, demonstrating the air stability of our Li ionic platform. As shown, the Raman maps exhibit strong intensities in the central opaque areas in Fig.4b. In contrast, the PL maps of the two flakes (shown in Fig. 4f) indicate intense optical emission on the edges of the flakes. The results clearly suggest that the AFM tip-induced Li intercalation effectively enhances the PL of multilayer MoS$_2$.

Figure 4g presents the PL spectra from representative locations marked by purple and light blue crosses in Fig. 4f. For clarity, the PL and Raman spectra from these regions are labeled as follows: purple-PL, purple-Raman, light blue-PL, light blue-Raman, and dark blue-Raman (Fig. 4d-h). Both PL spectra exhibit strong A peaks at 1.86 eV (purple) and 1.82 eV (light blue), respectively[4, 38].

For comparison, Fig. 4g also includes the PL spectrum of a CVD-grown monolayer $MoS_2$ film transferred onto LICGC (labeled CVD-PL). The CVD-PL peak intensity is stronger than that of the light blue-PL taken from inner flake region with lower Li intercalation. Strikingly, the purple-PL, taken from edge region with higher Li intercalation density, shows a peak intensity exceeding CVD-PL by 26-fold. This dramatic PL enhancement is a prominent evidence suggesting a transition from indirect to direct gap.

We propose that tip-induced Li intercalation decouples interlayer interactions in multilayer $MoS_2$, converting it into a vertically aligned stack of monolayers. This contrasts with prior reports, where complete interlayer decoupling in $MoS_2$ requires Li concentrations approaching 1 Li per unit cell—a regime that induces a metallic 1T phase and quenches PL entirely[24]. This difference likely arises from the distinct Li intercalation dynamics between solution-based methods and our approach.

To examine this claim, ultralow frequency (ULF) Raman spectroscopy was performed to characterize the locations indicated by the crosses in Fig. 4d, e. The ULF Raman characterizations are taken in a cross-circular polarization setup[39], focusing on the interlayer shear mode vibration that is highly sensitive to layer numbers[24, 37, 39]. The pink-Raman spectrum shown in Fig. 4h was obtained from the $MoS_2$ multilayer before AFM tip modulation. As shown, the shear mode of the pink-Raman spectrum peaks at 31.5 $cm^{-1}$, close to the value of bulk $MoS_2$. Upon Li intercalation, the inner regions of the left flake (light blue-Raman and dark blue-Raman) exhibit Raman peaks at wavenumbers below 31.5 $cm^{-1}$, suggesting the emergence of thinner layers due to Li intercalation even away from the edges. Notably, the dark blue - Raman spectrum, taken from the region closer to the intense PL area, shows a peak at 22.7 $cm^{-1}$, corresponding to bilayer $MoS_2$[37]. This Raman spectrum indicates the presence of a stack of various thin layers in this region. When moving to the region of intense PL, no shear mode is observed. Considering the appearance of intense PL, we can conclude that the intercalated Li ions effectively decouple the interlayer interactions of multilayer $MoS_2$ and transform it into a stack of multiple $MoS_2$ monolayers.

It is also of interest to note that the room temperature purple-PL is peaked at 1.86 eV, which is similar to that of mechanically-exfoliated monolayer $MoS_2$, indicating limited charge doping in the sample. In the Supplementary Fig. 10, we show low temperature PL where it is more straightforward to deconvolute the exciton and trion contributions. By comparing the exciton and trion energy and spectral weight, the Fermi energy of our $Li_xMoS_2$ is only about 15 meV[40]. These observations clearly indicate that the dynamics of AFM tip-induced Li intercalation are significantly different from those of solution-based Li intercalation[10, 21, 24, 26].

**Models for doping patterning via PFM Lithography**

Based on the above results, we can try to conclude the mechanism of patterning doping and Li intercalation using PFM lithography. First, the radius of the conductive AFM tip apex was observed using scanning electron microscopy (SEM), see Fig. 5a. Fig. 5b indicates the conductive AFM tip exhibiting nanoscale sharpness, with a radius of 25 nm[41]. According to basic electromagnetic principles, a high density of positive charges accumulates on the surface of the tip apex, generating a large and localized electric field when a positive voltage is applied to the AFM tip. This effect is widely utilized in field ion microscopy and atom probe tomography[42, 43]. When the tip voltage is fixed at $U$, the tip apex electric field $E$ is inversely proportional to the tip radius $R$, following the equation $E = \frac{U}{R}$.

We have established a simplified model: When the AFM tip contacts the upper surface of the dielectric material and a voltage $U$ is applied, a large electric field $E$ is generated due to the high density of charges accumulating on the tip's surface, a result of the tip enhancement effect. According to Gauss's law, the relationship between the electric field $E$ and the surface charge density $\sigma$ at the tip apex is given by the equation $E = \frac{\sigma}{4\pi\varepsilon}$, where $\varepsilon$ is the vacuum dielectric constant.

When the tip contacts the top surface of $Al_2O_3$, the dielectric film is polarized by the tip's electric field, inducing opposite charges on each side, see Fig. 5c. Here, we consider the scenario where the tip has a positive voltage and the contact region between the tip and $Al_2O_3$ is a nanoscale flat plane. Due to the formation of charges on the $Al_2O_3$ surface, an electric double layer (EDL) forms at the vicinal surface of the LICGC substrate because the Li ions move further into the bulk. When the AFM tip withdraws, the EDL in the LICGC remains, as the Li ions inside are static without external stimulation such as an electric field, as seen in Fig. 5d.

However, in the case of $Al_2O_3$/$MoS_2$/LICGC device with monolayer $MoS_2$, the EDL consists of a layer of negative charges in the $MoS_2$ and a complementary layer of positively charged Li ions in the LICGC, as shown in Fig. 5e. Upon withdrawing the AFM tip, the additional electron dopants remain in the monolayer $MoS_2$ (see Fig. 5f). Conversely, when a negative voltage is applied to the AFM tip, hole dopants are introduced into the monolayer $MoS_2$ to neutralize its native electron dopants. Concurrently, Li ions in the LICGC move away from the $MoS_2$, forming a negatively charged layer, see Supplementary Fig. 11. Consequently, the natively electron-doped $MoS_2$ exhibits increased resistance after the AFM tip is withdrawn, as illustrated in Supplementary Fig. 11b. The model explains the results presented in Fig. 2g-i.

Lastly, it should be noted that under a large positive tip voltage, Li intercalation occurs in the $Al_2O_3$/$MoS_2$/LICGC system with multilayer $MoS_2$, see Fig. 5g, h. This occurs because a large amount of negative charge can be generated in the multilayer $MoS_2$ thin flake when a large positive voltage is applied to the tip. Consequently, Li ions in the LICGC move towards $MoS_2$, leading to Li intercalation starting from the edges of the flake. This model can explain the results shown in Fig. 4. After withdrawing the AFM tip, the intercalated Li ions remain stationary. Alternatively, the LICGC substrate surface, with Li depletion, can provide a gating effect that partially neutralizes the electron doping induced by the intercalated Li ions, resulting in intense PL.

**Discussion**

In summary, we demonstrate a controllable ionic doping method by utilizing a solid Li-ion conductor in conjunction with a thin layer of an oxide dielectric film. The thin oxide film not only shields Li ions from the ambient environment, which is highly reactive to Li, but also permits the enhanced electric field of the AFM tip to control the motion of Li ions. The doping level and even Li intercalation can be regulated by the tip voltage, and the active region of Li-ion accumulation can be confined by the AFM nanotip head. Particularly, the tip-induced Li intercalation can completely screen the interlayer coupling of multilayer $MoS_2$, enabling the successful transformation of $MoS_2$ multilayer into vertically stacked multiple monolayers with a direct bandgap for strong light-matter interactions, without transitioning to the metallic 1T phase.

Our study confirms the localized high charge density enhanced by the nanotip can direct the forward and backward motion of the Li ions, depending on the voltage polarity, leading to the creation or erasure of doping patterns. The doping and ion intercalation technique presented herein is distinct from the existing methods[25,44,45], featuring high carrier density, customizable doping patterns, rewritability, air stability, and the nanoprobe fabrication of Li-intercalated materials with emerging phenomena. Consequently, our method is applicable to the development of various 2D electronics with strong light-matter interactions.

## Methods

### 2D materials preparation and transfer

Monolayer $MoS_2$ films were grown using the same CVD method reported[46]. Transfer of single-layer $MoS_2$ was performed using a PMMA-based method. Initially, the PMMA solution (950K A4, Micro Chem Corp) was spin-coated onto the single-layer $MoS_2$/sapphire substrate. The spin-coating was executed at a rate of 500 rpm for 5 seconds, followed by 2000 rpm for 55 seconds, and then dried for 3 minutes at a temperature of 80 °C. Subsequently, a potassium hydroxide (10 M, Sinopharm) solution was utilized to etch the sapphire substrate, allowing the PMMA/$MoS_2$ film to float on the solution's surface. The PMMA/$MoS_2$ film was then thoroughly washed with deionized water to eliminate any etchant residue. Following this, the PMMA/$MoS_2$ film was carefully scooped out using a LICGC substrate and dried overnight to remove any trapped water between the $MoS_2$ and the substrates. Lastly, the PMMA was removed by immersing the sample in an acetone bath and subsequently dried using compressed nitrogen gas.

### Device fabrication

Ultraviolet lithography (Suss MicroTec Lithography, MA/BA6 Gen4) was used to fabricate electrode patterns. Then, Ti/Au (2 nm/8 nm) was deposited by E-beam evaporation and lift-off with acetone. After the secondary ultraviolet lithography, $O_2$ plasmas were used to etch redundant single-layer $MoS_2$. Finally, the ALD system (SENTECH Instruments GmbH, SI ALD LL) was used to deposit $Al_2O_3$ with a thickness of about 27 nm at 150 °C.

### Raman and PL characterization

Raman and PL spectra were measured at room temperature using the Witec Alpha 300 RAS Raman spectrometer equipped with 300/1800 lines per mm gratings and a 532 nm wavelength laser. The power applied to the sample was approximately 3 mW for Raman and PL measurements. Scattered light was collected through a 100× objective (Zeiss LD EC Epiplan-Neofluar Dic 100× / 0.75). Raman and PL spectra measurement at low temperature using the Horiba and attocube equipped with 600/1800 lines per mm gratings and a 532 nm wavelength laser. The power applied to the sample was approximately 3 mW for Raman and PL measurements. Scattered light was collected through a 100× objective (LT-APO / 532-RAMAN / 0.82).

### AFM and electrical characterizations

The surface potential was modulated and analyzed via PFM and KPFM measurements with Cypher ES (Oxford) at room temperature. A Ti/Ir-coated conductive tip (Oxford, ASYELEC-01-R2) with a free resonance frequency of 75 kHz and a spring constant of 2.8 N m$^{-1}$ was used. In all PFM and KPFM measurements, the used scan rate is 1.0 Hz. The current-voltage characteristics were measured using Keysight B2901A SMU.

**Acknowledgements**
This work was supported by National Natural Science Foundation of China (Grant No. 12174319 to C.Z. and No. 12174318 to J.W.), Research Center for Industries of the Future (RCIF project No. WU2023C001 to J.W. and C.Z.) at Westlake University, the China Postdoctoral Science Foundation (Grant No. 2022M712844 to S.W.). J.Y. acknowledges support by the National Science Foundation (DMR-2004474) and is grateful for the hospitality of Westlake University. We thank L. Liu, from Instrumentation and Service Center for Physical Sciences, Z. Chen, and Y. Cheng from Instrumentation Service Center for Molecular Sciences at Westlake University, and we thank Westlake Center for Micro/Nano Fabrication for the facility support and technical assistance from X. Mu.

**Author contributions**
C.Z. conceived the project. Q.F. and Y.Z. carried out all the experiments with support from others. J.S. and W.K. prepared the $MoS_2$ films. J.Y. guided the Raman and PL measurements and assisted in the analysis of the relevant data. S.H. contributed to the setup of the low-temperature PL spectroscopy. G.Z. and J.W. fabricated the electrostatically gated exfoliated monolayer $MoS_2$ device. C.W. and J.S. provided assistance with the experimental details. C.Z., Q.F., Y.Z., J.W., and J.Y. analyzed the data and wrote the paper with input from all authors.

**Competing financial interests**
The authors declare no competing financial interests.

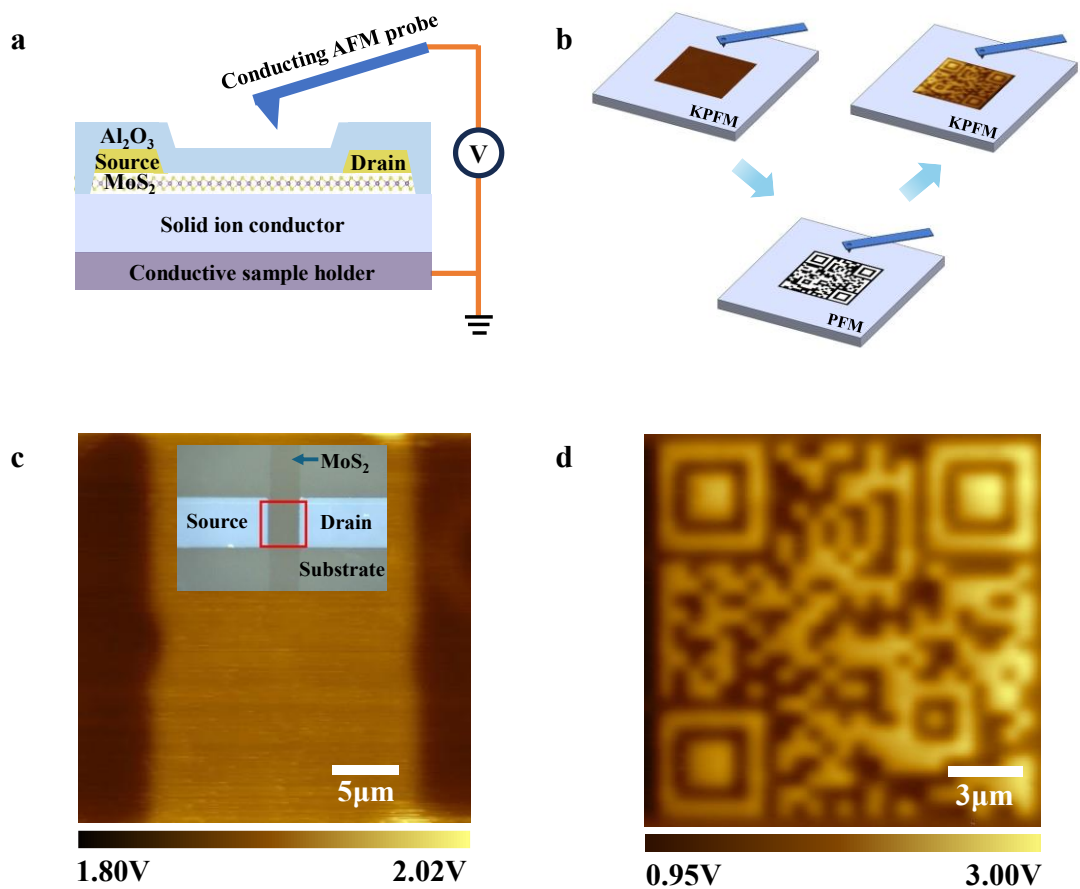

**Fig. 1: Concept of doping patterning electronics. a,** Schematic of the experimental setup. **b,** Fabrication steps for doping patterning using KPFM and piezoresponse force microscopy (PFM). **c,** KPFM image of an $Al_2O_3$/$MoS_2$/LICGC (Lithium-Ion Conducting Glass-Ceramics) device with a monolayer $MoS_2$ film. Inset: Optical image of the device, with the red box indicating the KPFM measurement region. **d,** KPFM image of a QR code doping pattern written via PFM lithography.

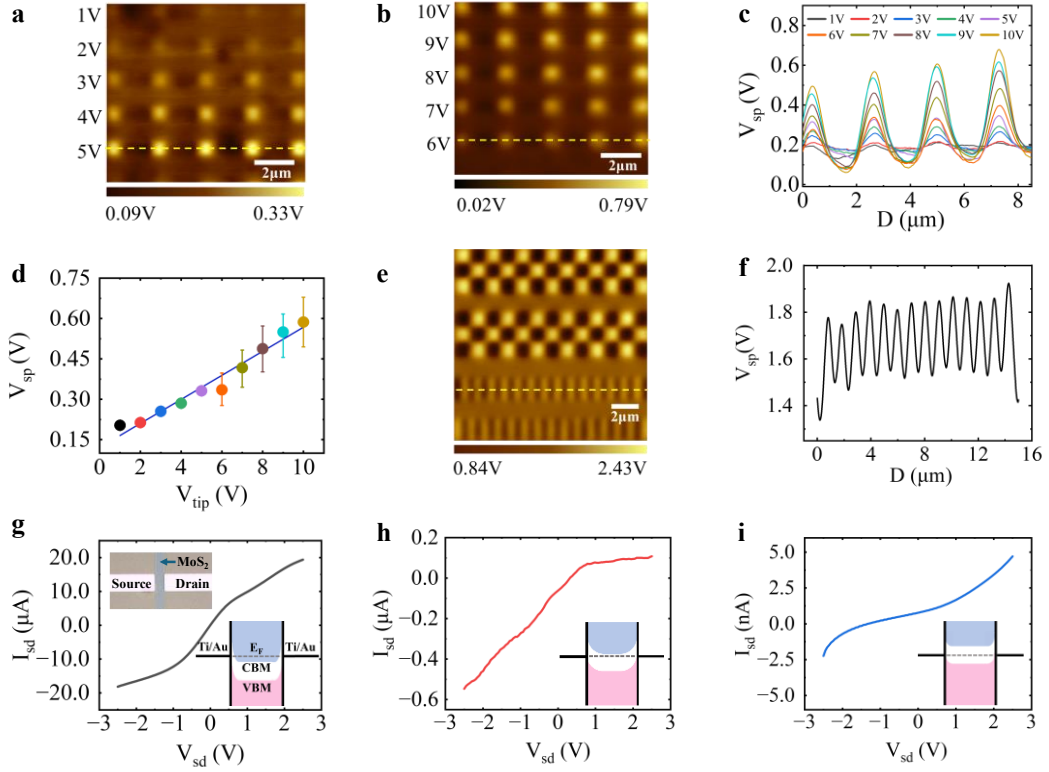

**Fig. 2: Modulations of doping patterns and device characteristics. a–b,** KPFM images of tip voltage series ranging from 1 V to 10 V in 1 V steps. **c,** Line profiles of surface potential ($V_{sp}$) patterns from (**a, b**). Yellow dashed lines in (**a, b**) mark typical extraction paths. **d,** Surface potential peak values (from **c**) versus PFM lithography voltage; blue line indicates linear fit. **e,** KPFM image of doping patterns written via PFM lithography at varying tip voltages. **f,** Surface potential line profile along the yellow dashed line in (**e**). Averaged full width at half maximum (FWHM) 517 nm defines doping resolution. **g–i,** Current–voltage (IV) evolution of an $Al_2O_3/MoS_2/LICGC$ device: (**g**) pristine state, (**h**) after −5 V PFM lithography, (**i**) after −8 V PFM lithography. Inset: Optical microscopy image of the device.

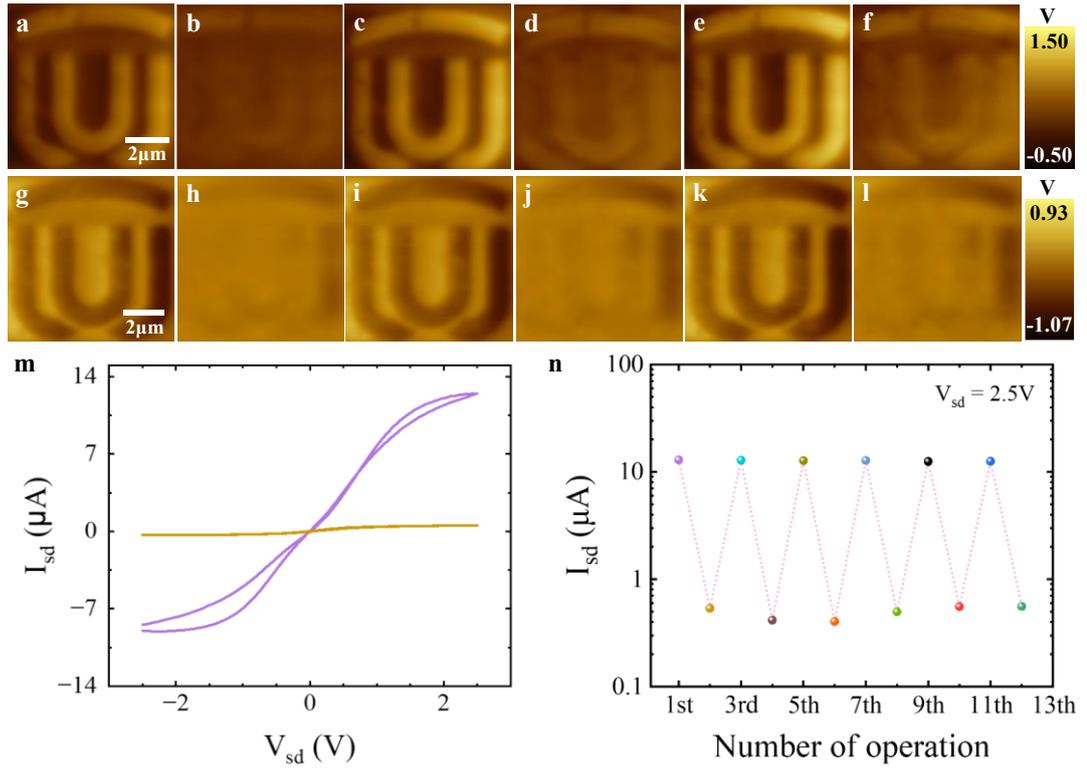

**Fig. 3: Erasable and repeatable ionic doping. a–f,** A doping pattern written with +8 V and erased with −3.4 V over three cycles. Scale bars in (**a**) applies to (**b–f**). **g–l,** The same doping pattern written with −8 V and erased with +2.5 V over three cycles. Scale bar in (**g**) applies to (**h–l**). **m,** $I$–$V$ curve measurements of an $Al_2O_3$/$MoS_2$/LICGC device, showing two resistance states switched by PFM lithography applied to the channel. **n,** Repeated source–drain current values ($I_{sd}$) at $V_{sd}$ = 2.5 V using PFM switching.

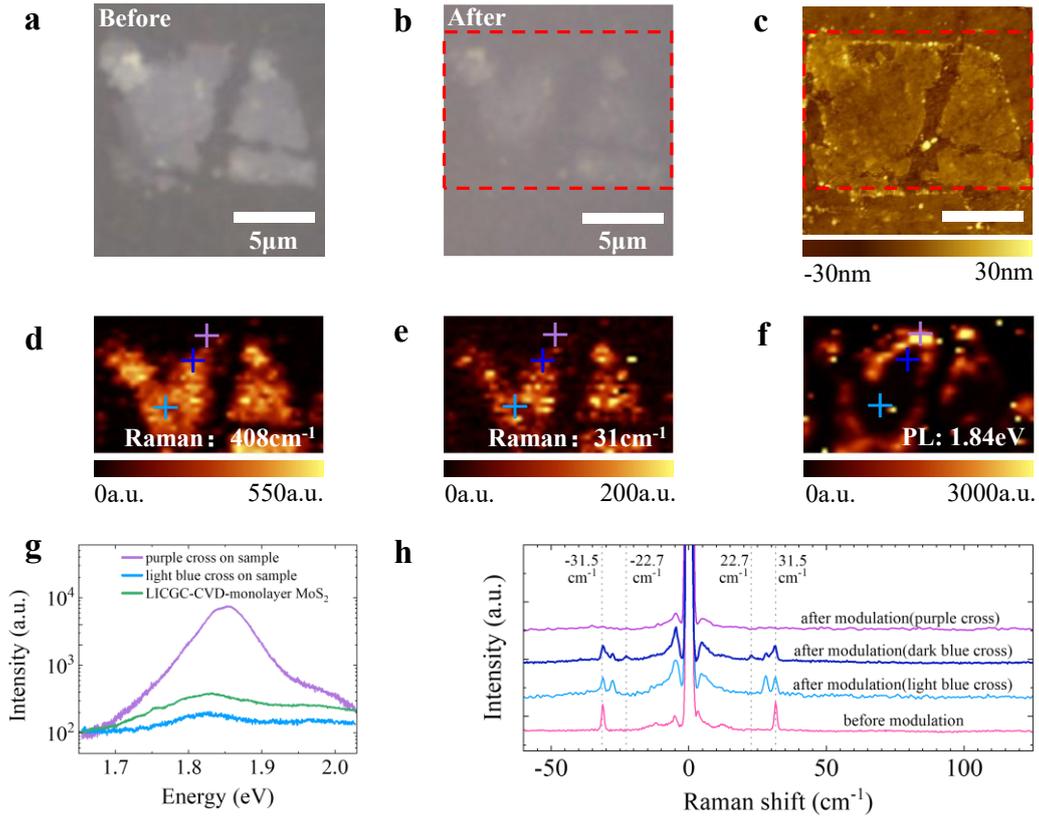

**Fig. 4: Impact of AFM tip-induced Li intercalation on optical properties of Li$_x$MoS$_2$. a, b,** Optical microscopy images of mechanically exfoliated MoS$_2$ flakes before (**a**) and after (**b**) Li intercalation via AFM tip modulation. **c,** AFM topography of the intercalated flakes (scale bar: 5 µm). **d–f,** Correlated Raman and photoluminescence (PL) mappings of the region highlighted in **b** and **c**. Crosses mark the locations for single-point spectral acquisition. **g**, PL spectra from the crosses in **f** and a chemical vapor deposition (CVD)-grown monolayer MoS$_2$ transferred onto a LICGC substrate. **h**, Ultralow-frequency (ULF) Raman spectra at the crosses in d, e, and for pristine multilayer MoS$_2$ flake (left flake in **a**).

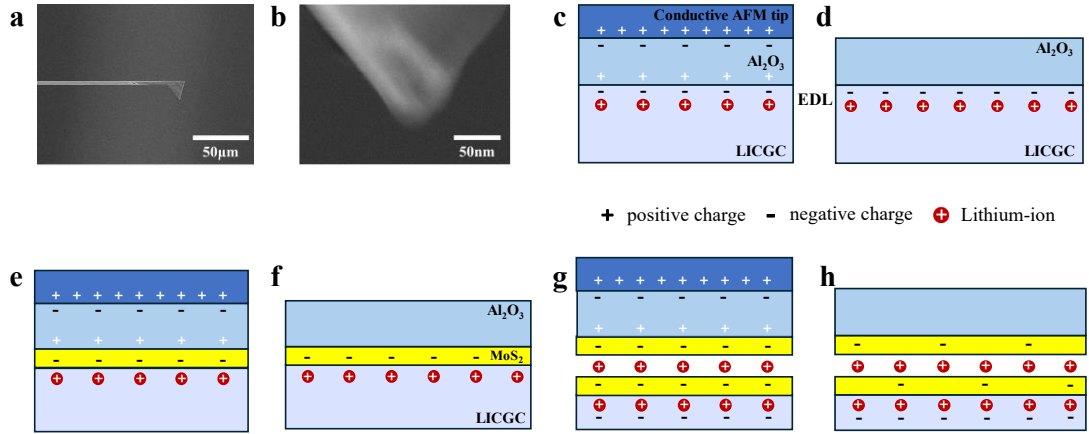

**Fig. 5: Mechanism of doping patterning and Li intercalation. a,** Scanning electron microscopy (SEM) image of an AFM cantilever. **b,** Zoomed-in SEM image of the AFM tip. **c,** Schematic of charge distribution during contact with an AFM tip at positive voltage during PFM lithography. **d,** Residual electric double layer (EDL) after tip withdrawal. **e,** Schematic of charge distribution in an $Al_2O_3$/$MoS_2$/LICGC device with monolayer $MoS_2$ upon contact with a positively biased tip. **f,** Residual EDL after tip withdrawal. **g,** Schematic of an $Al_2O_3$/$MoS_2$/LICGC device with Li intercalation and bilayer/multilayer $MoS_2$ upon contact with a positively biased tip. The negative charges at the bottom are caused by Li depletion. **h,** Residual intercalated Li ions after tip withdrawal, illustrating the combined doping effects of intercalated Li and the LICGC substrate with Li depletion layer on $MoS_2$ layers.

# Supplementary Information

Supplementary Fig. 1. Optical spectra measurements.
Supplementary Fig. 2. Calibration of AFM tip surface potential using an Au-Si-Al standard sample.
Supplementary Fig. 3. PFM lithography pattern and surface morphology.
Supplementary Fig. 4. Comparison of bare LICGC surface before and after PFM lithography.
Supplementary Fig. 5. Evolutions of the surface potentials of the two vicinal regions with and without PFM lithography, respectively.
Supplementary Fig. 6. Voltage maps of chess board and bars for PFM lithography.
Supplementary Fig. 7. Air stability of Li doped $MoS_2$ electronics.
Supplementary Fig. 8. Voltage maps of a logo for PFM lithography.
Supplementary Fig. 9. Height profile of the $MoS_2$ flake after Li intercalation.
Supplementary Fig. 10. Low temperature (2K) PL.
Supplementary Fig. 11. Schematic of PFM lithography with negative voltage.

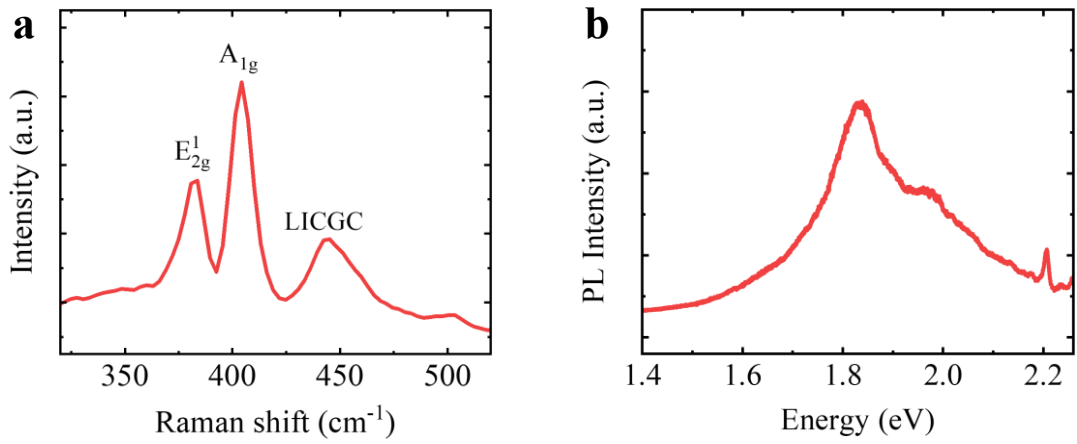

**Supplementary Fig. 1. Optical spectra measurements. a,** Raman spectrum of monolayer $MoS_2$ channel of a device. **b,** Photoluminescence spectrum of the monolayer $MoS_2$ channel.

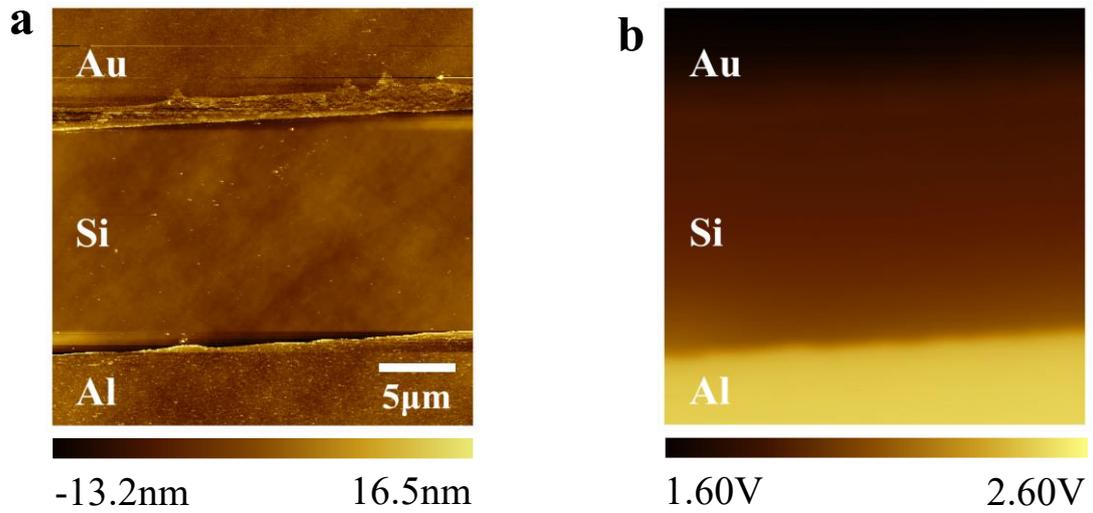

**Supplementary Fig. 2. Calibration of AFM tip surface potential using an Au-Si-Al standard sample. a,** AFM topography of a Au-Si-Al calibration sample. **b,** The corresponding KPFM image of the Au-Si-Al standard sample. The reference value of the fresh Au surface potential is 5.1 eV. According to the average surface potential of the Si and Al regions shown in (**b**), their calculated values are 4.815 eV and 4.295 eV, respectively.

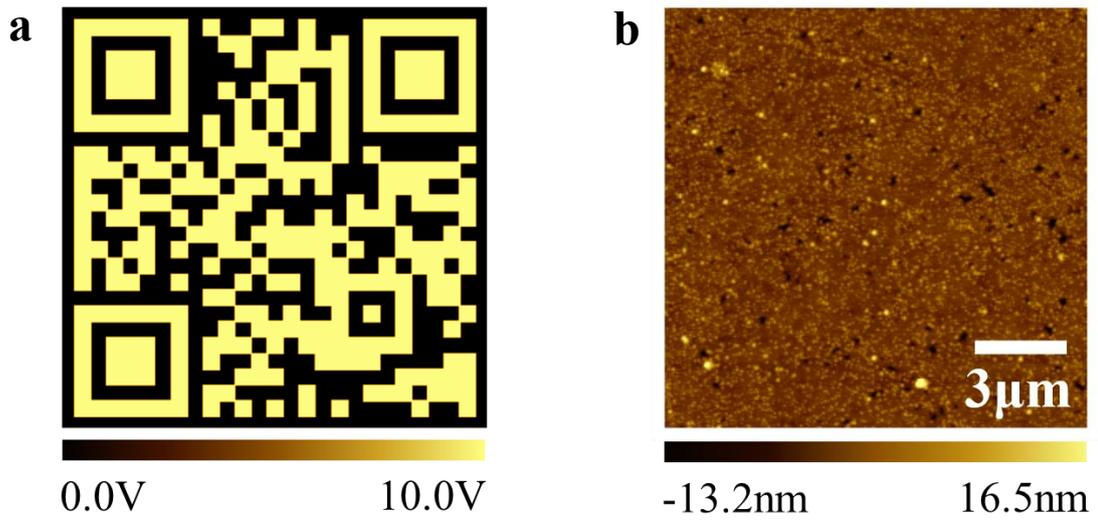

**Supplementary Fig. 3. PFM lithography pattern and surface morphology. a,** The tip voltage map of a quick response (QR) code. **b,** Corresponding AFM topography of the region on $Al_2O_3$/LICGC after PFM lithography. The change of surface morphology after PFM lithography does not happen.

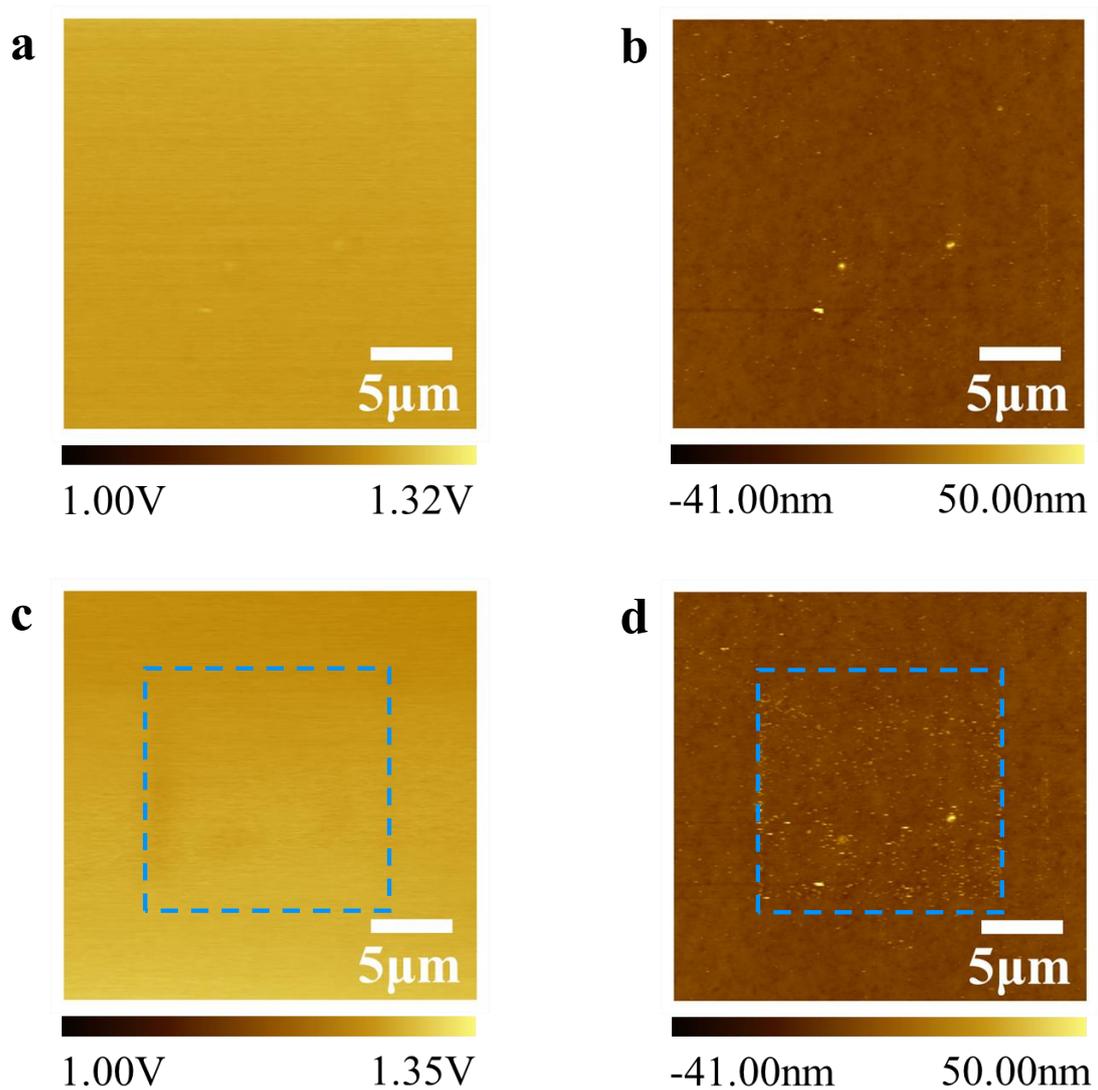

**Supplementary Fig. 4. Comparison of bare LICGC surface before and after PFM lithography. a,** KPFM image of the bare LICGC surface before PFM lithography. **b,** Corresponding AFM topography of the region. **c,** KPFM image of the same region after PFM lithography. The dashed blue square indicates the region written by 5 V PFM lithography. No surface potential difference is observed. **d,** Corresponding AFM topography of the same surface. The topography of the region after PFM lithography remains similar except some small dots emerge.

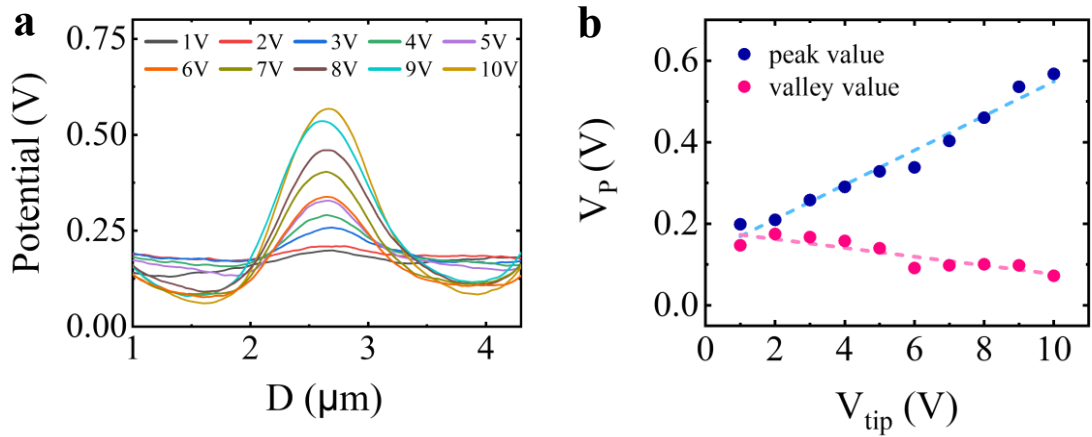

**Supplementary Fig. 5. Evolutions of the surface potentials of the two vicinal regions with and without PFM lithography, respectively. a,** Surface potential curves of a single PFM lithography pattern and the regions nearby. The curves are extracted from **Fig. 2c**. b, Evolution of peak value and valley value as a function of tip voltage. The peak values and valley values are taken from the peaks and the left valleys shown in **a**.

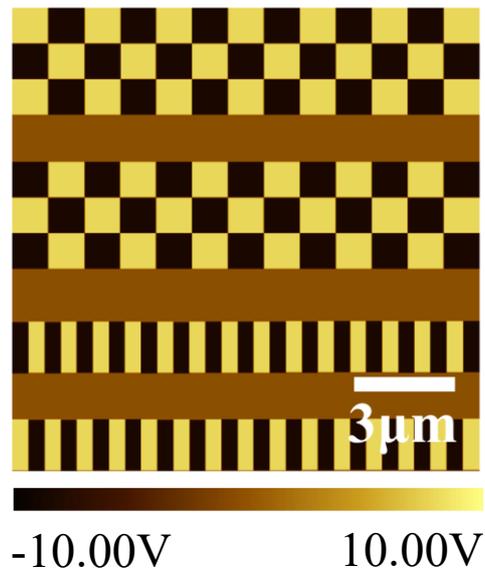

**Supplementary Fig. 6. Voltage maps of chess board and bars for PFM lithography.**

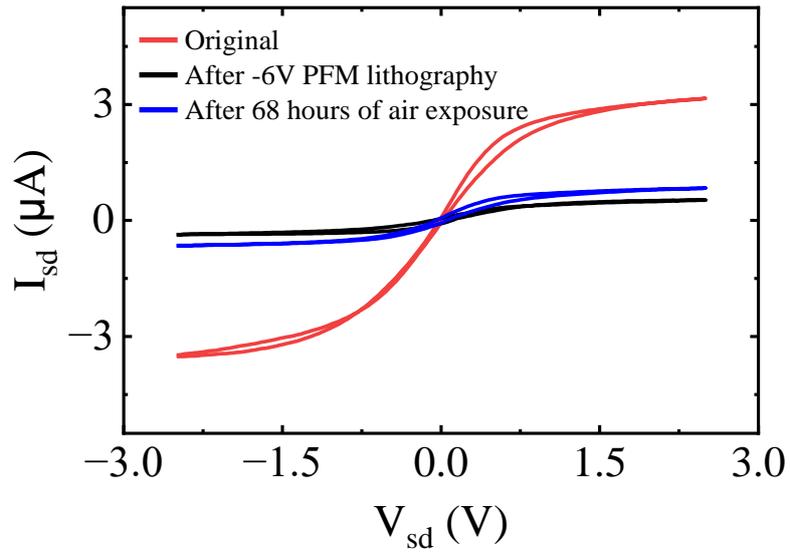

**Supplementary Fig. 7. Air stability of Li doped MoS$_2$ electronics.** I-V curves taken from an Al$_2$O$_3$/1L-MoS$_2$/LICGC device at three different stages in a series of original, after -6 V PFM lithography, and after 68 hours of air exposure.

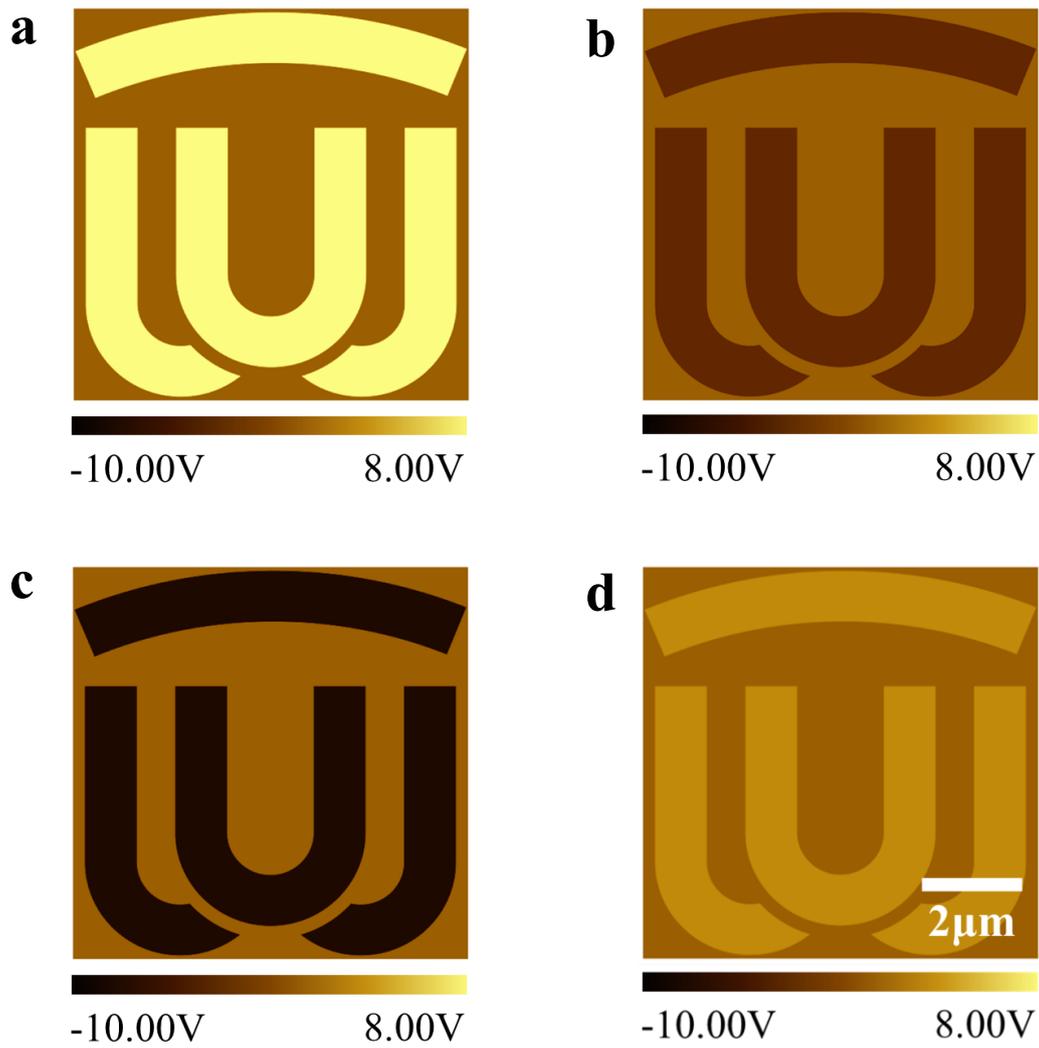

**Supplementary Fig. 8. Voltage maps of a logo for PFM lithography. a,** Positive voltage map for writing dopant. **b,** Negative voltage map for erasing dopant. **c,** Negative voltage map for writing dopant. **d,** Positive voltage map for erasing dopant.

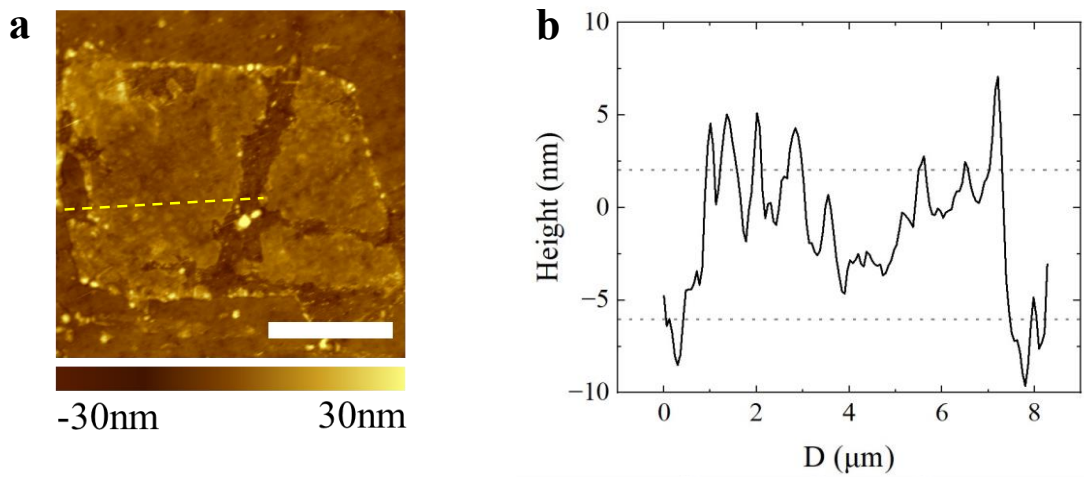

**Supplementary Fig. 9. Height profile of the MoS$_2$ flake after Li intercalation. a,** AFM tomography of Li intercalated MoS$_2$ flake indicating a dash line for taking height profile (scale bar: 5 µm). **b,** Height profile of the Li intercalated MoS$_2$ flake.

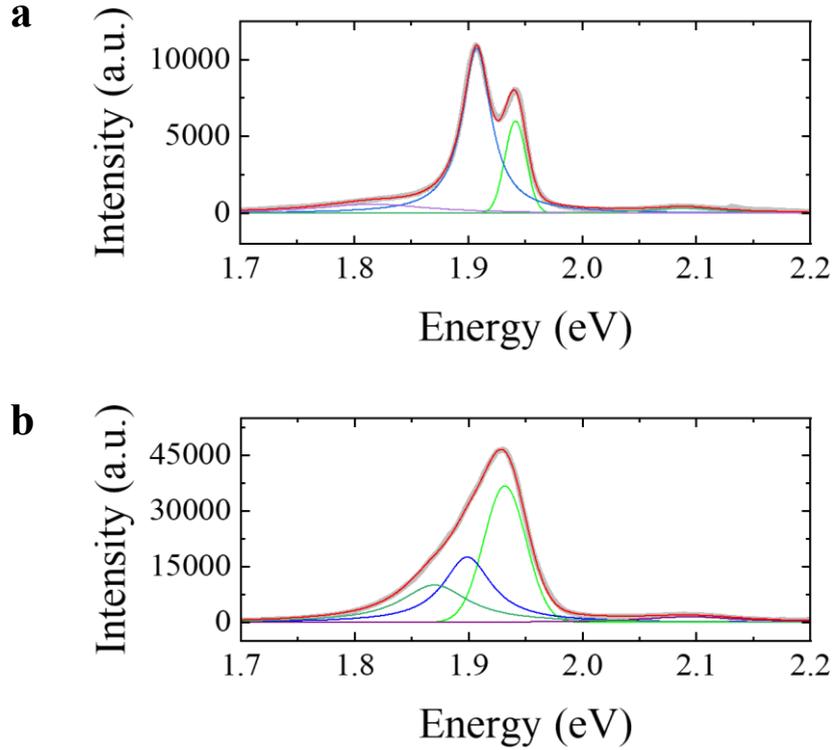

**Supplementary Fig. 10. Low temperature (2K) PL. a,** From an exfoliated monolayer $MoS_2$ sample sandwiched between hexagonal boron nitride on $Si/SiO_2$. **b,** From Li interacted $Al_2O_3/MoS_2/LICGC$ device. Thick gray traces are experimental data, and smooth colored curves are multi-peak fittings. The green components are due to neutral excitons, and the blue are due to trions. The 2.1 eV bump is from the B exciton, and the lowest energy peak arises from disorder-trapped excitons. For the as-made monolayer $MoS_2$, the spectra weight of trion is higher than that of exciton, while for the Li modulated multilayer $MoS_2$, the two spectral weights are comparable. The exciton-trion splitting is 34 meV for (**a**), and 33 meV for (**b**). Both observations indicate that the electron doping level of our Li intercalated device is not more than the as-made monolayer $MoS_2$. From the trion binding energy of 18 meV, we deduce that the Fermi energy in our direct gap device is about 15 meV.

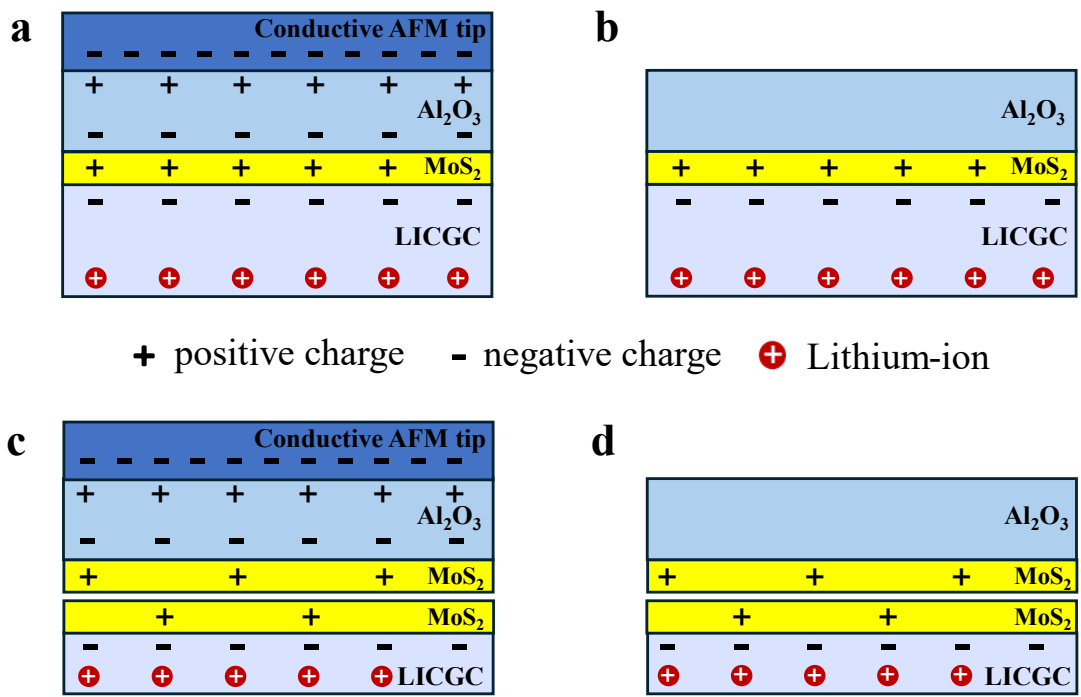

**Supplementary Fig. 11. Schematic of PFM lithography with negative voltage. a, b,** Models for Al$_2$O$_3$/MoS$_2$/LICGC device with monolayer MoS$_2$. **c, d,** Models for Al$_2$O$_3$/MoS$_2$/LICGC device with multilayer MoS$_2$.